\documentclass{article}
\usepackage{amsmath}

\input{tcilatex}
\begin{document}

\begin{titlepage}
\begin{flushright}
\end{flushright}
\vspace*{3cm}
\begin{center}
{\Large \textsf{\textbf{Anisotropic universe space-time non-commutativity and scalar particle creation in the presence of a constant electric field}}}
\end{center}
\par \vskip 5mm
\begin{center}
{\large \textsf{Slimane Zaim}}\\\vskip
5mm
D\'{e}partement  des Sciences de la Mati\`{e}re, Facult\'{e} des Sciences,\\
Universit\'{e} Hadj Lakhdar - Batna, Algeria. \\
\end{center}
\par \vskip 2mm
\begin{center} {\large \textsf{\textbf{Abstract}}}\end{center}
\begin{quote}
We study the effect of the non-commutativity on the creation of scalar particles from vacuum in the anisotropic universe space-time.
We derive the deformed Klein-Gordon equation up to second order in the non-commutativity parameter using the general modified field equation. Then the canonical method based on Bogoliubov transformation is applied to calculate the  probability of particle creation in vacuum and  the corresponding number density in the $k$ mode. We deduce that the non-commutative space-time introduces a new source of particle creation.
\end{quote} \vspace*{2cm}


\noindent\textbf{\sc Pacs numbers}: 11.10.Nx, 03.65.Pm, 03.70.+k , 25.75.Dw

\end{titlepage}

\section{Introduction}

In the classical theory black holes can only absorb and not emit particles.
However it was shown that quantum mechanical effects cause black holes to
create and emit particles. It is also well known that the most significant
prediction of this theory is the phenomenon of particle creation which leads
to the concept of quantum gravity. In this paper we are interested in the
issue of particle production by constructing a simple type of the
non-commutative geometry.

The extension of quantum field theory to one in a curved space is the
starting point towards quantum gravity in curved space-time. However another
important concept in the context of quantum gravity is non-commutative
geometry, by which the quantization of the space-time leads to quantifying
gravity. Thus the non-commutative space-time is intrinsically connected to
quantum gravity despite the well-known problem of Lorentz-violating
symmetry. All other fundamental problems, such as the unitarity violation $%
\left[ 4\right] $, causality $\left[ 5\right] $ and UV/IR divergences $\left[
6\right] $, have been discussed in the context of the local Lorentz
invariance.

In ref. $[7]$, the authors showed that these problem can arise by inducing a
non-constant metric into the theory and they found that at high energy
gravity and non-commutative geometry must become dependant on each other.
Several important works were performed in a formally Lorentz-invariant
approach (see for example the reviews $\left[ 2,8,9\right] $). Various
theories of gravity in the context of non-commutative geometry have amongst
others been studied in refs. $[10-22]$ and cosmology on non-commutative
space-time has been explored in ref. $[23]$. Even certain ideas referring to
quantum gravity have been explored with respect to non-commutative geometry,
see refs. $[24-28]$ for example. Another approach is based on the twisted
Poincar\'{e} algebra constructed for canonically deformed space with a
constant parameter of non-commutativity, where in this formalism the
Lagrangian density is invariant and the gauge and pure gravity theories are
consistent.

In our previous work $[29]$ we have attempted to construct a non-commutative
gauge gravity model, where the problem of the unitarity (see for example
refs. $[4,,14,15,30,31,32]$) is overcome by the construction of generalized
local Lorentz and general coordinate transformations, which preserve the
non-commutative coordinate canonical commutation relations. The phenomenon
of scalar particle creation in anisotropic Bianchi I universe with a
constant electric field has been analyzed in ref. $\left[ 33\right] $.
Actually, there is no electric charge in the universe to create an electric
field, meaning that the particles can be created from vacuum by the
expansion of the universe itself with no other external field present.

The aim of this paper is the study the effect of the non-commutativity on
the creation of scalar particles from vacuum in the space-time anisotropic
Bianchi I universe when a constant electric field is present. We compute the
number density of created particles in the cases of strong and weak field.
From our results we clearly deduce that the the non-commutativity plays the
role of the electric field.

This paper is organized as follows. In section 2 we derive the corresponding
Seiberg-Witten maps up to the first order of $\theta $ for the various
dynamical fields and we propose an invariant action of the pure
non-commutative gauge gravity and non-commutative charged scalar field in
interaction. In section 3 we derive the anisotropic universe
non-commutativity space-time Klein-Gordon (KG) equation and obtain its
solution. Then we compute the density of created scalar particles and
discuss the weak and strong field limits. The last section is devoted to a
discussion.

\section{Seiberg-Witten maps}

One can get at first order in the non-commutative parameter $\theta ^{\mu
\nu }$ the following Seiberg--Witten maps $\left[ 1\right] $: 
\begin{align}
\hat{\varphi}& =\varphi -\frac{1}{2}\theta ^{\mu \nu }A_{\nu }\partial _{\mu
}\varphi +\mathcal{O}\left( \theta ^{2}\right) ,  \notag \\
\hat{\lambda}_{P}& =\lambda _{P}+\frac{1}{4}\theta ^{\sigma \rho }\left\{
\partial _{\sigma }\lambda _{P},\omega _{\rho }\right\} \mathcal{+O}\left(
\theta ^{2}\right) ,  \notag \\
\hat{\lambda}_{G}& =\lambda _{G}+\frac{1}{4}\theta ^{\sigma \rho }\left\{
\partial _{\sigma }\lambda _{G},A_{\rho }\right\} +\mathcal{O}\left( \theta
^{2}\right) , \\
\hat{A}_{\xi }& =A_{\xi }-\frac{1}{4}\theta ^{\mu \nu }\left\{ A_{\nu
},\partial _{\mu }A_{\xi }+F_{\mu \xi }\right\} +\mathcal{O}\left( \theta
^{2}\right) ,  \notag \\
F_{\mu \xi }^{1}& =\frac{1}{2}\theta ^{\alpha \beta }\left\{ F_{\mu \alpha
}F_{\xi \beta }\right\} -\frac{1}{4}\theta ^{\alpha \beta }\left\{ A_{\alpha
},\left( \partial _{\beta }+D_{\beta }\right) F_{\mu \xi }\right\} +\mathcal{%
O}\left( \theta ^{2}\right) ,  \notag \\
\hat{e}_{\mu }^{a}& =e_{\mu }^{a}-\frac{i}{4}\theta ^{\alpha \beta }\left(
\omega _{\alpha }^{ac}\partial _{\beta }e_{\mu }^{c}+\left( \partial _{\beta
}\omega _{\mu }^{ac}+R_{\beta \mu }^{ac}\right) e_{\mu }^{c}\right) +%
\mathcal{O}\left( \theta ^{2}\right) ,  \notag
\end{align}
where 
\begin{align}
F_{\mu \nu }& =\partial _{\mu }A_{\nu }-\partial _{\nu }A_{\mu }-i\left[
A_{\mu },A_{\nu }\right] , \\
\omega _{\mu }& =\omega _{\mu }^{ab}S_{ab}, \\
\hat{A}_{\mu }& =\hat{A}_{\mu }^{a}T^{a}=\hat{A}_{k}^{a}\ast \hat{e}_{\mu
}^{k}T^{a}, \\
\hat{\omega}_{\mu }& =\hat{\omega}_{\mu }^{ab}S_{ab}=\hat{\omega}%
_{k}^{ab}\ast \hat{e}_{\mu }^{k}S_{ab}, \\
\theta ^{\mu \nu }& =\hat{e}_{\ast a}^{\mu }\ast \hat{e}_{\ast b}^{\mu
}\theta ^{ab},
\end{align}%
and $\omega _{\mu }^{ab}$ are the spin connections and $\hat{e}_{\ast
a}^{\mu }$ is the inverse-$\ast $ of the vierbein $\hat{e}_{\mu }^{a}$
defined as: 
\begin{equation}
\hat{e}_{\mu }^{b}\ast \hat{e}_{\ast a}^{\mu }=\delta _{a}^{b},
\end{equation}%
and 
\begin{equation}
\hat{e}_{\mu }^{a}\ast \hat{e}_{\ast a}^{\nu }=\delta _{\mu }^{\nu }.
\end{equation}

To begin we consider a non-commutative gauge theory with a charged scalar
particle in the presence of an electrodynamic gauge field in a general
curvilinear system of coordinates. We can write the action as: 
\begin{equation}
\mathcal{S}=\frac{1}{2\kappa ^{2}}\int d^{4}x\text{ }(\mathcal{L}_{G}+%
\mathcal{L}_{SC})\,,  \label{eq:action}
\end{equation}%
where $\mathcal{L}_{G}$ and $\mathcal{L}_{SC}$ stand for the pure gravity
and matter scalar densities corresponding to the charged scalar particle in
the presence of an electric field, and where 
\begin{equation}
\mathcal{L}_{G}=\hat{e}\ast \hat{R}\,,
\end{equation}%
and 
\begin{equation}
\mathcal{L}_{SC}=\hat{e}\ast \left( \hat{g}^{\mu \nu }\ast \left( \hat{D}%
_{\mu }\hat{\varphi}\right) ^{\dagger }\ast \hat{D}_{\nu }\hat{\varphi}+m^{2}%
\hat{\varphi}^{\dagger }\ast \hat{\varphi}\right)\, .
\end{equation}%
The deformed tetrad and scalar curvature are given by: 
\begin{align}
\hat{e}& =\det {}_{\ast }(\hat{e}_{\mu }^{a})\equiv \frac{1}{4!}\epsilon
^{\mu \nu \rho \sigma }\varepsilon _{abcd}\hat{e}_{\mu }^{a}\ast \hat{e}%
_{\nu }^{b}\ast \hat{e}_{\rho }^{c}\ast \hat{e}_{\sigma }^{d}\,, \\
\hat{R}& =\hat{e}_{\ast a}^{\mu }\ast \hat{e}_{\ast b}^{\nu }\ast \hat{R}%
_{\mu \nu }^{ab}\,,
\end{align}%
and the gauge covariant derivative is defined as: $\hat{D}_{\mu }\hat{%
\varphi }=\left( \partial _{\mu }-ie\hat{A}_{\mu }\right) \ast \hat{\varphi}$%
.

In the following we consider a symmetric metric $\hat{g}_{\mu \nu }$ such
that: 
\begin{equation}
\hat{g}_{\mu \nu }=\frac{1}{2}(\hat{e}_{\mu }^{b}\ast \hat{e}_{\nu b}+\hat{e}%
_{\nu }^{b}\ast \hat{e}_{\mu b}).
\end{equation}%
As a consequence, the first-order expansion in the non-commutative parameter 
$\theta ^{\alpha \beta }$ of the scalar curvature $\hat{R}$ and metric $\hat{%
g}_{\mu \nu }$ vanishes. Thus $\hat{R}$ and $\hat{g}_{\mu \nu }$ can be
rewritten as: 
\begin{align}
\hat{R}& =R+\mathcal{O}\left( \theta ^{2}\right) , \\
\hat{g}_{\mu \nu }& =g_{\mu \nu }+\mathcal{O}\left( \theta ^{2}\right) ,
\end{align}

Next we use the generic field infinitesimal transformations ($\hat{\delta}_{%
\hat{\lambda}}\hat{\varphi}=i\hat{\lambda}\ast \hat{\varphi}$), and the
star-product tensor relations. We can prove that the action in eq. $\left(
34\right) $ is actually invariant. By varying the scalar density under the
gauge transformation and from the generalised field equation and the Noether
theorem we obtain $\left[ 10\right] $: 
\begin{equation}
\frac{\partial \mathcal{L}}{\partial \hat{\varphi}}-\partial _{\mu }\frac{%
\partial \mathcal{L}}{\partial \left( \partial _{\mu }\hat{\varphi}\right) }%
+\partial _{\mu }\partial _{\nu }\frac{\partial \mathcal{L}}{\partial \left(
\partial _{\mu }\partial _{\nu }\hat{\varphi}\right) }+\mathcal{O}\left(
\theta ^{2}\right) =0.  \label{eq:field}
\end{equation}

\section{The solution to the non-commutative Klein-Gordon equation and
particle creation process}

In this section we examine the particle creation phenomenon induced by
vacuum instabilities in the context of the non-commutative geometry in
presence the external vector potential $A_{\mu }$. We shall take the example
of the Klein-Gordon equation in a cosmological anisotropic non-commutative
Bianchi $I$ universe.

The deformed line element of the Bianchi $I$ universe up to the first-order
of $\theta $ takes the following form: 
\begin{equation}
ds^{2}=-dt^{2}+t^{2}\left( dx^{2}+dy^{2}\right) +dz^{2}+g_{\mu \nu }^{\left(
1\right) }dx^{\mu }dx^{\nu }+\mathcal{O}\left( \theta ^{2}\right) .
\label{eq:metric}
\end{equation}%
We choose for $\theta ^{\alpha \beta }$ the following form: 
\begin{equation}
\theta ^{\alpha \beta }=\left( 
\begin{array}{cccc}
0 & 0 & 0 & \theta \\ 
0 & 0 & \theta & 0 \\ 
0 & -\theta & 0 & 0 \\ 
-\theta & 0 & 0 & 0%
\end{array}%
\right) ,\qquad \alpha ,\,\beta =0,1,2,3.
\end{equation}%
We follow the same steps outlined in ref. $\left[ 33\right] $ and look for
the non-commutative correction of the metric up to the first order in $%
\theta $.

Choosing the following diagonal tetrads: 
\begin{align}
\underline{e}_{\mu }^{0}& =\left( 
\begin{array}{cccc}
1, & 0, & 0, & 0%
\end{array}%
\right) , \\
\underline{e}_{\mu }^{1}& =\left( 
\begin{array}{cccc}
0, & t, & 0, & 0%
\end{array}%
\right) , \\
\underline{e}_{\mu }^{2}& =\left( 
\begin{array}{cccc}
0, & 0, & t, & 0%
\end{array}%
\right) , \\
\underline{e}_{\mu }^{3}& =\left( 
\begin{array}{cccc}
0, & 0, & 0, & 1%
\end{array}%
\right) ,
\end{align}%
then the nonzero spin connections are 
\begin{align}
\omega _{1}^{01}& =-\omega _{1}^{10}=1, \\
\omega _{2}^{02}& =-\omega _{2}^{20}=1.
\end{align}%
Using the Seiberg-Witten map $(1)$ and the choice $(19)$ we can obtain the
following deformed veirbeins: 
\begin{align}
\underline{\hat{e}}_{\mu }^{0}& =\left( 
\begin{array}{cccc}
1, & 0, & 0, & 0%
\end{array}%
\right) , \\
\underline{\hat{e}}_{\mu }^{1}& =\left( 
\begin{array}{cccc}
0, & t, & -i\frac{\theta }{4}t, & 0%
\end{array}%
\right) , \\
\underline{\hat{e}}_{\mu }^{2}& =\left( 
\begin{array}{cccc}
0, & i\frac{\theta }{4}t, & t, & 0%
\end{array}%
\right) , \\
\underline{\hat{e}}_{\mu }^{3}& =\left( 
\begin{array}{cccc}
0, & 0, & 0, & 1%
\end{array}%
\right) .
\end{align}
As a consequence, the first-order expansion in the non-commutative parameter 
$\theta ^{\alpha \beta }$ of the Bianchi $I$ metric vanishes. Thus (18) can
be rewritten as:%
\begin{equation}
ds^{2}=-dt^{2}+t^{2}\left( dx^{2}+dy^{2}\right) +dz^{2}+\mathcal{O}\left(
\theta ^{2}\right) .
\end{equation}

In order to identify the particle states we follow the quasi-classical
approach of ref. $\left[ 34\right] $. The standard method is to specify the
positive and negative frequency modes and solve the classical
Hamilton-Jacobi equation looking specifically for the asymptotic limits of
the solution $t\rightarrow 0$ and $t\rightarrow \infty $. Then one solves
the Klein Gordon equation by comparing with the quasi-classical limits, and
specifying the positive and negative frequency states. Finally one utilises
Bogouliubov transformations and calculates the number density for created
particles.

Using the modified field equation $\left( 17\right) $ with the generic boson
field $\hat{\varphi}$ one can find in a non-commutative curved space-time
and in the presence of the external potential $\hat{A}_{\mu }$ the following
modified Klein-Gordon equation: 
\begin{equation}
\left( \eta ^{\mu \nu }\partial _{\mu }\partial _{\nu }-m_{e}^{2}\right) 
\hat{\varphi}\,+\left( ie\eta ^{\mu \nu }\partial _{\mu }\hat{A}_{\nu
}-e^{2}\eta ^{\mu \nu }\hat{A}_{\mu }\ast \hat{A}_{\nu }+2ie\eta ^{\mu \nu }%
\hat{A}_{\mu }\partial _{\nu }\right) \hat{\varphi}=0,
\end{equation}
with the deformed external potential $\hat{A}_{\mu }=\left( 0,0,0,Et\right) $
in free non-commutative space-time being: 
\begin{equation}
\hat{a}_{3}=a_{3}-\Theta ^{\mu k}a_{k}\partial _{\mu }a_{3}+\mathcal{O}%
\left( \Theta ^{2}\right) .
\end{equation}%
For a non-commutative time-space we have $\Theta ^{03}\neq 0$ and $\Theta
^{ki}=0$, where $i,k=1,2,3$. In this case we can write: 
\begin{equation}
\eta ^{\mu \nu }\partial _{\mu }\partial _{\nu }=-\partial _{0}^{2}+\frac{1}{%
t^{2}}\left( \partial _{1}^{2}+\partial _{2}^{2}\right) +\partial _{3}^{2},
\end{equation}%
and 
\begin{equation}
2ie\eta ^{\mu \nu }\hat{A}_{\mu }\partial _{\nu }=2ieEt\left( 1+\theta
E\right) \partial _{3},
\end{equation}%
and 
\begin{equation}
-e^{2}\eta ^{\mu \nu }\hat{A}_{\mu }\ast \hat{A}_{\nu }=\left[ ieEt\left(
1+\theta E\right) \right] ^{2}\,.
\end{equation}

The Klein-Gordon equation $\left( 31\right) $ (in the presence of a constant
external field $A_{\mu }$) up to $\mathcal{O}\left( \theta ^{2}\right) $
then simplifies to: 
\begin{equation}
\left[ -\partial _{0}^{2}+\frac{1}{t^{2}}\left( \partial _{1}^{2}+\partial
_{2}^{2}\right) +\partial _{3}^{2}-m^{2}+2ieEt\left( 1+\theta E\right)
\partial _{3}+\left[ ieEt\left( 1+\theta E\right) \right] ^{2}\right] \hat{%
\varphi}=0.
\end{equation}
In order to keep our results compact and transparent we make use of the
approximation: 
\begin{equation}
1+\theta g\approx \exp \left( \theta g\right),
\end{equation}
with $g$ being an arbitrary regular function. Equation $\left( 36\right) $
commutes with the operator $-i\overrightarrow{\nabla }$, and therefore the
wave functions $\hat{\varphi}$ can be cast into: 
\begin{equation}
\hat{\varphi}=\tilde{\Delta}(t)\exp \left( ik_{x}x+ik_{y}y+ik_{z}z\right).
\end{equation}
Substituting eq. $\left( 38\right) $ into eq.$\left( 36\right) $, one can
get: 
\begin{equation}
\left[ \frac{d^{2}}{dt^{2}}+\frac{k_{\bot }^{2}}{t^{2}}+k_{z}^{2}+m^{2}+2e%
\tilde{E}tk_{z}+e^{2}\tilde{E}^{2}t^{2}\right] \tilde{\Delta}(t)=0\,,
\end{equation}
where 
\begin{equation}
\tilde{E}=E\exp \left( \theta E\right),
\end{equation}
and the eigenvalue $k_{\bot }$ is given by: 
\begin{equation}
k_{\bot }=\sqrt{k_{x}^{2}+k_{y}^{2}}.
\end{equation}

We adopt the following change of variable: 
\begin{equation}
\rho =ie\tilde{E}t^{2},
\end{equation}%
and we deduce that for $k_{z}=0,$ equation $\left( 39\right) $ becomes: 
\begin{equation}
\left[ \frac{d^{2}}{d\rho ^{2}}+\frac{1}{2\rho }\frac{d}{d\rho }+\frac{%
k_{\bot }^{2}}{4\rho }-\frac{1}{4}-i\frac{m^{2}}{4e\tilde{E}}\right] \tilde{%
\Delta}(\rho )=0.
\end{equation}

Following ref. $\left[ 34\right] $, the solution to eq. $\left( 43\right) $
can be written as a combination of Whittaker functions $M_{\tilde{k}_{\theta
},\mu }\left( \rho \right) $ and $W_{\tilde{k}_{\theta },\mu }\left( \rho
\right) $: 
\begin{equation}
\tilde{\Delta}\left( \rho \right) =\rho ^{-1/4}\left( C_{1}M_{\tilde{k}%
_{\theta },\mu }\left( z\right) +C_{2}W_{\tilde{k}_{\theta },\mu }\left(
z\right) \right) ,  \label{eq:sol}
\end{equation}
where $\tilde{k}_{\theta }$ and $\mu $ are given by: 
\begin{equation}
\tilde{k}_{\theta }=-i\frac{m^{2}}{4eE}\exp \left( -\theta E\right) ,\qquad
\mu =\frac{i}{2}\sqrt{k_{\bot }^{2}-\frac{1}{4}}\,.
\end{equation}
Then the general solution of $(43)$ can be expressed in terms of the
hypergeometric functions $F\left( \frac{1}{2}-\tilde{k}_{\theta }+\mu ,2\mu
+1,\rho \right) $ and $G\left( \frac{1}{2}-\tilde{k}_{\theta }+\mu ,2\mu
+1,\rho \right) $ as follows: 
\begin{align}
\tilde{\Delta}\left( \rho \right) = & C_{1}\rho ^{\mu +1/4}e^{-\rho
/2}F\left( \frac{1}{2}-\tilde{k}_{\theta }+\mu ,2\mu +1,\rho \right) + 
\notag \\
&+C_{2}\rho ^{\mu+1/4}e^{-\rho /2}G\left( \frac{1}{2}-\tilde{k}_{\theta
}+\mu ,2\mu +1,\rho \right) ,
\end{align}
where $C_{1}$ and $C_{2}$ are normalisation constants.

To construct the positive and negative frequency modes we use the asymptotic
limit of the solution $\left( 46\right) $ and compare the result with that
obtained by solving the Hamilton-Jacobi relativistic equation at $t=0$ $%
\left( \rho =0\right) $. Thus it may be shown that the positive and negative
frequency modes are given by: 
\begin{equation}
\tilde{\Delta}_{0}^{+}=C_{0}^{+}\rho ^{\mu +1/4}e^{-\rho /2}F\left( \frac{1}{%
2}-\tilde{k}_{\theta }+\mu ,2\mu +1,\rho \right) ,
\end{equation}%
and 
\begin{equation}
\tilde{\Delta}_{0}^{-}=\left( \tilde{\Delta}_{0}^{+}\right) ^{\ast
}=C_{0}^{+}\left( -1\right) ^{\mu +\frac{1}{4}}\rho ^{\mu +1/4}e^{-\rho
/2}F\left( \frac{1}{2}-\tilde{k}_{\theta }+\mu ,2\mu +1,\rho \right) ,
\end{equation}%
where $C_{0}^{+}$ is normalisation constant. We note that the hypergeometric
functions $F\left( \frac{1}{2}-\tilde{k}_{\theta }+\mu ,2\mu +1,\rho \right) 
$ and $G\left( \frac{1}{2}-\tilde{k}_{\theta }+\mu ,2\mu +1,\rho \right) $
have the following asymptotic limits: 
\begin{align}
F\left( \frac{1}{2}-\tilde{k}_{\theta }+\mu ,2\mu +1,\rho \right) &\sim 1
\qquad & \text{for}& \qquad \left\vert \rho \right\vert \ll 1, \\
G\left( \frac{1}{2}-\tilde{k}_{\theta }+\mu ,2\mu +1,\rho \right) &\sim \rho
^{\tilde{k}_{\theta }-\mu -1/2} \qquad & \text{for} &\qquad \left\vert \rho
\right\vert \rightarrow \infty .
\end{align}

One can show that the positive and negative frequency modes for $\left\vert
\rho \right\vert \rightarrow \infty $, by observing the asymptotic limit of $%
G\left( \frac{1}{2}-\tilde{k}_{\theta }+\mu ,2\mu +1,\rho \right) $, is
given by: 
\begin{equation}
\tilde{\Delta}_{\infty }^{+}=C_{\infty }^{+}\rho ^{\mu +1/4}e^{-\rho
/2}G\left( \frac{1}{2}-\tilde{k}_{\theta }+\mu ,2\mu +1,\rho \right) ,
\end{equation}%
and 
\begin{equation}
\tilde{\Delta}_{\infty }^{-}=C_{\infty }^{-}\left( -\rho \right) ^{\mu
+1/4}e^{\rho /2}G\left( \frac{1}{2}+\tilde{k}_{\theta }+\mu ,2\mu +1,-\rho
\right) ,
\end{equation}%
where $C_{\infty }^{+}$ and $C_{\infty }^{-}$ are normalisation constants.

Now we utilise the relation: 
\begin{multline}
G\left( \frac{1}{2}-\tilde{k}_{\theta }+\mu ,2\mu +1,\rho \right) =\frac{%
\Gamma \left( -2\mu \right) }{\Gamma \left( \frac{1}{2}-\tilde{k}_{\theta
}-\mu \right) }F\left( \frac{1}{2}-\tilde{k}_{\theta }+\mu ,2\mu +1,\rho
\right) + \\
+\frac{\Gamma \left( 2\mu \right) }{\Gamma \left( \frac{1}{2}-\tilde{k}%
_{\theta }+\mu \right) }\rho ^{-2\mu }F\left( \frac{1}{2}-\tilde{k}_{\theta
}-\mu ,-2\mu +1,\rho \right) ,
\end{multline}%
where $\Gamma $ is the Gamma function, and exploit the fact that the
positive frequency mode $\tilde{\Delta}_{\infty }^{+}$ can be written in
terms of the positive ($\tilde{\Delta}_{0}^{+}$) and negative ($\tilde{\Delta%
}_{0}^{-}$) frequency modes through the Bogouliubov transformation $\left[
35,36,37\right] $: 
\begin{equation}
\tilde{\Delta}_{\infty }^{+}=\hat{\alpha}\tilde{\Delta}_{0}^{+}+\hat{\beta}%
\tilde{\Delta}_{0}^{-},  \label{eq:positive}
\end{equation}%
to find that $\hat{\alpha}$ and $\hat{\beta}$ are: 
\begin{equation}
\hat{\alpha}=\frac{C_{\infty }^{+}\Gamma \left( -2\mu \right) }{%
C_{0}^{+}\Gamma \left( \frac{1}{2}-\tilde{k}_{\theta }-\mu \right) },\qquad 
\hat{\beta}=\frac{C_{\infty }^{+}\Gamma \left( 2\mu \right) }{%
C_{0}^{+}\Gamma \left( \frac{1}{2}-\tilde{k}_{\theta }+\mu \right) }\exp
\left( i\pi \left( \mu +1/4\right) \right) ,
\end{equation}%
with 
\begin{equation}
\frac{\left\vert \hat{\alpha}\right\vert ^{2}}{\left\vert \hat{\beta}%
\right\vert ^{2}}=\left\vert \frac{\Gamma \left( \frac{1}{2}-\tilde{k}%
_{\theta }+\mu \right) }{\Gamma \left( \frac{1}{2}-\tilde{k}_{\theta }-\mu
\right) }\right\vert ^{2}\exp \left( 2\pi \mu \right) .
\end{equation}

Using the following property of the Gamma function: 
\begin{equation}
\left\vert \Gamma \left( \frac{1}{2}+i\rho \right) \right\vert ^{2}=\frac{%
\pi }{\cosh \left( \pi \rho \right) },
\end{equation}
and simplifying leads to: 
\begin{equation}
\frac{\left\vert \hat{\alpha}\right\vert ^{2}}{\left\vert \hat{\beta}%
\right\vert ^{2}}=\frac{\cosh \left[ \pi \left( -\tilde{k}_{\theta }+\mu
\right) \right] }{\cosh \left[ \pi \left( \tilde{k}_{\theta }+\mu \right) %
\right] }\exp \left( 2\pi \mu \right).
\end{equation}
The probability to create a single particle from vacuum is then: 
\begin{equation}
P_{k}=\left( \frac{\left\vert \hat{\alpha}\right\vert ^{2}}{\left\vert \hat{%
\beta}\right\vert ^{2}}\right) ^{-1}=\frac{\cosh \left[ \pi \left( \tilde{k}%
_{\theta }+\mu \right) \right] }{\cosh \left[ \pi \left( -\tilde{k}_{\theta
}+\mu \right) \right] }\exp \left( -2\pi \mu \right) .
\end{equation}
Taking into account the fact that $\tilde{k}_{\theta }=\frac{m^{2}}{4eE}%
-\theta \frac{m^{2}}{4e},$ for small $\theta $, we easily to obtain: 
\begin{equation}
P_{k}=P_{k}\left( \theta =0\right) +P_{k}^{\theta }\,,
\end{equation}
where $P_{k}\left( \theta =0\right) $ denotes the ordinary probability to
create a single particle from vacuum in the presence of an electric field
and has the expression: 
\begin{equation}
P_{k}\left( \theta =0\right) =\frac{\cosh \left[ \pi \left( k+\mu \right) %
\right] }{\cosh \left[ \pi \left( -k+\mu \right) \right] }\exp \left( -2\pi
\mu \right) ,\text{ \ }k=\frac{m^{2}}{4eE}\,,
\end{equation}
and $P_{k}^{\theta }$ is the generated non-commutative correction of order $%
\theta $ given by: 
\begin{equation}
P_{k}^{\theta }=-\frac{\pi m^{2}}{4e}\theta P_{k}\left( \theta =0\right) %
\left[ \tanh \pi \left( k+\mu \right) +\tanh \pi \left( -k+\mu \right) %
\right].
\end{equation}

Next we calculate the non-commutative density of the created particles $\hat{%
n}$ by the non-commutative curved space-time and electric field. For this we
use eq. $\left( 54\right) $ so as to arrive at: 
\begin{equation}
\hat{n}=\left\vert \hat{\beta}\right\vert ^{2}\,.
\end{equation}
Using the normalisation condition $\left[ 37\right]$: 
\begin{equation}
\text{ }\left\vert \hat{\alpha}\right\vert ^{2}-\left\vert \hat{\beta}%
\right\vert ^{2}=1\,,
\end{equation}%
we finally arrive at the result for the non-commutative number density of
created particles $\hat{n}$: 
\begin{equation}
\hat{n}=\left( \frac{1-P_{k}}{P_{k}}\right) ^{-1}=\exp \left[ \pi \left( 
\tilde{k}_{\theta }-\mu \right) \right] \frac{\cosh \left[ \pi \left( \tilde{%
k}_{\theta }+\mu \right) \right] }{\sinh \left( 2\pi \mu \right) }\,.
\end{equation}

It is also very important to consider the weak and strong electric field
limits and see the behavior of the number density and derive some of the
related thermodynamical quantities.

\subsection{The weak field approximation}

In this limit, if we set: 
\begin{eqnarray}
\tilde{k}_{\theta } &=&\frac{m^{2}}{4eE}\left( 1-\theta E\right)  \notag \\
&=&\frac{m^{2}}{4eE}-\theta \frac{m^{2}}{4e}\,,
\end{eqnarray}
such that: 
\begin{equation}
\tilde{k}_{\theta }\rightarrow \infty\,,
\end{equation}
it is easy to show that the probability $P_{k}$ takes the form: 
\begin{equation}
P_{k}=\exp \left[ -\pi \left( \sqrt{k_{\bot }^{2}-\frac{1}{4}}+\theta \frac{%
m^{2}}{4e}\right) \right].
\end{equation}
Then the number density $\hat{n}$ is written up to the second order of $%
\theta $ as: 
\begin{equation}
\hat{n}=\frac{1}{\exp \left[ \pi \left( \sqrt{k_{\bot }^{2}-\frac{1}{4}}%
+\theta \frac{m^{2}}{4e}\right) \right] -1}\,.
\end{equation}

This density is thermal and looks like a two-dimensional Bose-Einstein
distribution with chemical potential $\mu ^{\theta }=-\theta \frac{\pi m^{2}%
}{4e}$. To get the total non-commutative number of the created particles per
a unit volume, we have to integrate the density $\hat{n}$ over momentum
space. Taking into account the fact that $\hat{n}$ does not explicitly
depend on $k_{z}$, the total non-commutative number $\hat{N}$ reads: 
\begin{equation}
\hat{N}=\frac{2}{\left( 2\pi T\right) ^{2}}\int \hat{n}k_{\bot }dk_{\bot
}dk_{z}\,,
\end{equation}%
where $T$ is the time for the external interaction and the integration over $%
k_{z}$ is equivalent to the integration of the classical equation of motion: 
$\int dk=\int eEdt=eET$. Thus the total non-commutative number $\hat{N}$ per
a unit volume takes the form: 
\begin{equation}
\hat{N}=\frac{2eET}{\left( 2\pi T\right) ^{2}}\left[ \int \frac{k_{\bot
}dk_{\bot }}{e^{\pi \left( \sqrt{k_{\bot }^{2}-\frac{1}{4}}+\theta \frac{%
m^{2}}{4e}\right) }-1}\right].
\end{equation}
Now since $\theta $ is small we have $\exp \left( \theta \frac{m^{2}}{4e}%
\right) \ll 1$. Consequently the total number $\hat{N}$ in eq.$(65)$,
written up to the second order of $\theta $, is given by the following
relation: 
\begin{equation}
\hat{N}\cong \frac{eE}{2\pi ^{4}T}\exp \left( -\pi \frac{m^{2}}{4e}\theta
\right) \left( 1+\frac{1}{4}\exp \left( -\pi \frac{m^{2}}{4e}\theta \right)
\right) +\mathcal{O}\left( \theta ^{2}\right).
\end{equation}

Notice that the particle creation mechanism is effectively isotropic in the
presence of a constant electric field of the anisotropic Bianchi I universe
of the non-commutative space-time. The non-commutative number density of
created particles in eq.$\left( 66\right) $ takes a similar form in the
Boltzmann limit in ordinary commutative space with a chemical potential $\mu
^{\theta }$ $=-\theta \frac{m^{2}}{4e}$. This result was expected due to the
fact that the non-commutativity parameter is the smallest area in space that
can be probed.

\subsection{The Strong Field Approximation}

In this limit if we set 
\begin{eqnarray}
\tilde{k}_{\theta } &=&\frac{m^{2}}{4eE}\left( 1-\theta E\right)  \notag \\
&=&k-\theta \frac{m^{2}}{4e}\,,
\end{eqnarray}
such that: 
\begin{equation}
k\rightarrow 0\,,
\end{equation}
and by direct simplifications one may show that the number density $\hat{n}$
takes the form: 
\begin{equation}
\hat{n}=\frac{1}{\exp \left[ \pi \left( \sqrt{k_{\bot }^{2}-\frac{1}{4}}%
+\theta \frac{m^{2}}{2e}\tanh \pi \sqrt{k_{\bot }^{2}-\frac{1}{4}}\right) %
\right] -1}\,.
\end{equation}%
This result looks like a two dimensional Bose-Einstein distribution with a
chemical potential $\mu^{\theta }$ given by: 
\begin{equation}
\mu ^{\theta }=-\theta \frac{\pi m^{2}}{2e}\tanh \pi \sqrt{k_{\bot }^{2}-%
\frac{1}{4}}\,.
\end{equation}

Consequently the results shown in eqs. $(69)$ and $(75)$ indicate that the
particle creation mechanism effectively isotropizes in the non-commutative
space-time of an anisotropic Bianchi I universe, where the number density of
created particles in the absence of the electric field corresponds to a
thermal distribution. Here the non-commutativity parameter plays the role of
electric field. In the limit $\theta \rightarrow 0$ equations $(69)$ and $%
(75)$ reduce to: 
\begin{equation*}
\frac{1}{\exp \left[ \pi \left( \sqrt{k_{\bot }^{2}-\frac{1}{4}}\right) %
\right] -1}\,,
\end{equation*}%
which is the same as the Bose-Einstein distribution of particles ( see $[38]$%
).

\section{Conclusions}

In this work we started with the effect of the non-commutativity on the
creation of scalar particles from vacuum in the space-time anisotropic
Bianchi I universe. By using the Seiberg-Witten maps and the Moyal product
up to first order in the non-commutativity parameter $\theta $, we
generalised the deformed Klein Gorden equation. After solving this equation
the density of created particles are calculated by applying the Bogoliubov
transformations and the quasi-classical limit for identifying the positive
and negative frequency modes. We have seen that the non-commutative
space-time introduces a new source of particle creation by considering the
application of our findings into the Bianchi I universe. As a conclusion,
the non-commutativity plays the same role as the electric field and chemical
potential. It is worth mentioning that in the limit $\theta \rightarrow
E^{-1}$ (i.e. the case of strong field) and at high energies our results
coincide with those of ref.$[33]$.

\section*{Acknowledgement}

I am grateful to Yazid Delenda for useful discussions and suggestions. This
work is supported by the CNEPRU project: D01320130009.

\end{document}